\documentclass[a4paper,12pt]{article}
\usepackage{amssymb}
\usepackage{graphicx}

\def\ie{\textit{i.e. }}
\def\pd{\partial}

\def\l{{\cal L}}

\def\u{\vec{v}}
\def\a{\vec{A}}

\def\pd{\partial}

\def\be{\begin{equation}}
\def\ee{\end{equation}}
\def\bea{\begin{eqnarray}}
\def\eea{\end{eqnarray}}

\def\({\left(}
\def\){\right)}
\def\[{\left[}
\def\]{\right]}

\def\j#1#2#3#4{{\it #1} {\bf #2} #3 #4}

\title{Large amplitude of the internal motion of DNA immersed in bio-fluid}

\author{A. Sulaiman\footnote{Email : lyman@tisda.org}}
\date{}

\begin{document}

\maketitle

\thispagestyle{empty}

\begin{center}
\begin{small}
\noindent
Geostech BPPT\footnote{http://www.bppt.go.id}, Kompleks Puspiptek Serpong, Tangerang 15310, Indonesia \\
\end{small}
\end{center}

\vspace*{5mm}

\begin{abstract}
The interaction between large internal motion of DNA surrounded by bio-fluid is investigated. The phenomenon is modelled using the relativistic Navier-Stokes lagrangian describing the bio-fluid coupled to the standard Klein-Gordon lagrangian describing the DNA. It is shown that the equation of motion at non-relativistic limit, $|\u| \ll c$, reproduces the well-known Sine-Gordon equation. The effect of the interaction in a single soliton solution is also given and discussed.
\end{abstract}

\vspace*{5mm}

\clearpage

From physical point of view, DNA is considered as a system consisting of many interacting molecules in a particular configuration of space-time. It has been shown that under particular external conditions the molecules form a double helix \cite{yakushevich}. The helix has a dynamic and flexible structure. The motion (transverse, longitudinal and torsional) of DNA can be divided in two main regions : the small and large amplitude of internal motions. The small amplitude of motion can be described by the hamiltonian of harmonic oscillator. On the other hand, the large amplitude is described by a non-harmonic one \cite{mingalev, radha}. 

Recently, many works have discussed and arrived at the conclusion that the large amplitude of internal motion can be considered as a nonlinear dynamical system where solitary conformational waves can be excited \cite{yakushevich}. Nonlinear interaction between molecules in DNA gives rise to a very stable excitation, the so-called soliton \cite{mingalev}. Soliton is a pulse-like nonlinear wave which forms a collision with similar pulse having static shape and speed \cite{scott}.

As mentioned above, DNA is not motionless. It is in a constantly wringgling dynamics state in a medium of bio-organic fluid in the nucleus cell. However, the motion of DNA surrounded by fluid is rarely studied. Previous studies are usually done by solving the Navier-Stokes equation and its wave equation simultaneously using appropriate boundary conditions. On the other hand, in the Hamiltonian formulation the viscous force is considered to be comparable with other forces arising from Hamiltonian \cite{kovici}. The solution is then obtained by expansion and performing order-by-order calculation. In these approaches, anyway the picture of interaction between DNA and its surrounding fluid is not clear.

Now, in the present paper a new approach to investigate the interaction between DNA and bio-fluid is discussed using the lagrangian method. Rather putting it by hand, the interaction is described in more natural way from the first principle, \ie by introducing some symmetries in the lagrangian under consideration. This kind of Navier-Stokes lagrangian has successfully been developed for both relativistic and non-relativistics cases \cite{leman2, leman1, leman3}. Imposing an appropriate (gauge) symmetry to the bosonic lagrangian with boson field $\Phi$, one can construct a gauge invariant lagrangian. The lagrangian induces gauge bosons $A_\mu$ as follows \cite{leman3},  
\be
 	\l_\mathrm{NS} = \left( \pd_\mu \Phi \right)^\dagger  \left( \pd^\mu \Phi \right) + V(\Phi) + \l_A \; , 
	\label{eq:l}
\ee
where, 
\be
	\l_A = -\frac{1}{4} F^a_{\mu\nu} {F^a}^{\mu\nu} + g J^a_\mu {A^a}^\mu + \frac{i}{2} f^{abc} g^2 \left( \Phi^\dagger T^a \Phi \right) A_\mu^b {A^c}^\mu \; , 
	\label{eq:la}
\ee
the strength tensor is $F^a_{\mu\nu} \equiv \pd_\mu A^a_\nu - \pd_\nu A^a_\mu - g f^{abc} A^b_\mu A^c_\nu$, while the 4-vector current is,
\be
	J^a_\mu = [ (\pd_\mu \Phi)^\dagger T^a \Phi - \Phi^\dagger T^a (\pd_\mu \Phi)] \; , 
	\label{eq:j}
\ee
and satisfies the current conservation $\pd^\mu J^a_\mu = 0$ respectively. The additional terms in $\l_A$ are required to keep invariances of the bosonic lagrangian under local (non-Abelian) gauge transformation $U \equiv \mathrm{exp}[-i T^a \theta^a(x)]$  \cite{chengli}, where $T_a$'s are generators belong to a particular Lie group and satisfy certain commutation  relation $[T^a,T^b] = i f^{abc} T^c$ with $f^{abc}$ is the anti-symmetric structure constant. It has further been shown that the relativistic Navier-Stokes equation can be reproduced for \cite{leman3}, 
\be
A_\mu = \left( \Phi, \a \right) \equiv \left( -c^2\sqrt{1-{|\u|^2}/{c^2}}, -\u \right) \; . 
	\label{eq:a}
\ee
While at non-relativistic limit, $v \ll c$, it coincides to the classical Navier-Stokes equation.

\begin{figure}[t]
  	\centering \includegraphics[width=10cm]{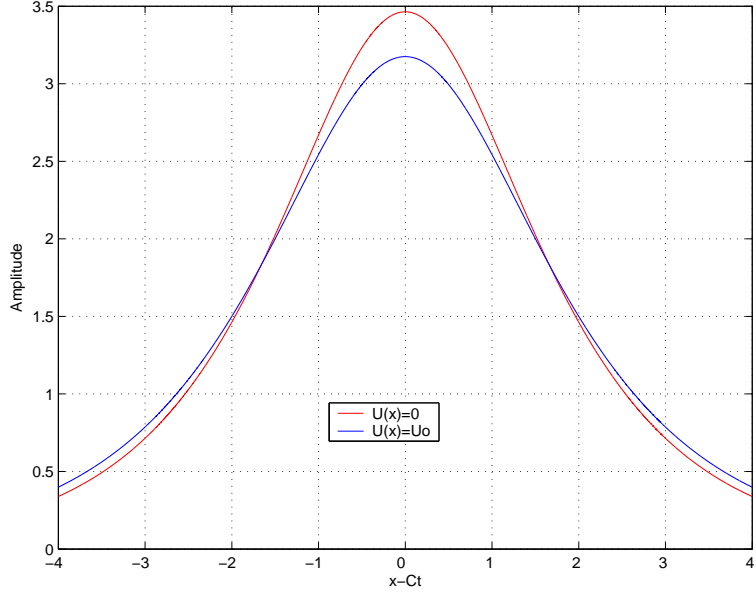}
  	\caption{Single soliton solution of the nonlinear Klein-Gordon equation. }
	\label{fig:soliton}
\end{figure}

Torsional motion of DNA in the continuous approximation can be reproduced by taking the potential $V(\Phi)$ that is familiar in the $\Phi^4-$theory \cite{chengli}, 
\be
	V(\Phi) = \frac{m^2}{2} \Phi^2 - \frac{\lambda}{4!} \Phi^4 \; ,
	\label{eq:v}
\ee
where $m$ and $\lambda$ are constants, and the dimensions are $[m] = 1$ and $[\lambda] = 0$ in the mass unit. Especially $\lambda$ determines the 'level of non-linearity' for the Klein-Gordon equation. Using the Euler-lagrange principle, Eqs. (\ref{eq:l}), (\ref{eq:a}) and (\ref{eq:v}) yield the equation of motion in term of $\Phi$, 
\be
 \frac{\pd^2 \Phi}{\pd t^2} - \frac{\pd^2 \Phi}{\pd x^2}
 - (m^2 - g^2 U(x)^2) \Phi + g^2 \left( \frac{\pd U(x)}{\pd x} \right)^2 \Phi +
 \frac{\lambda}{3!} \Phi^3 = 0 \; ,
 \label{eq:sg4}
\ee
 at non-relativistic limit for laminar flow, \ie $\u = U(x) \hat{i}$, in two-dimensional space $(t, x)$. In a special case when $U(x) \equiv U$ is a constant, Eq. (\ref{eq:sg4}) reads, 
\be
 \frac{\pd^2 \Phi}{\pd t^2} - \frac{\pd^2 \Phi}{\pd x^2}
 - (m^2 - g^2 U^2) \Phi + \frac{\lambda}{3!} \Phi^3=0 \; ,
 \label{eq:sg2}
\ee 
that is the non-linear Klein-Gordon equation for $\lambda \neq 0$. Further, if $m^2 - g^2 U^2 \approx \lambda$,
\be
 \frac{\pd^2 \Phi}{\pd t^2} - \frac{\pd^2 \Phi}{\pd x^2}  - \lambda \sin \Phi = 0 \; ,
 \label{eq:sg1}
\ee
using Taylor expansion : $\sin \Phi \approx \Phi - \frac{1}{3!} \Phi^3 + \cdots$. This is nothing else than the Sine-Gordon equation \cite{yakushevich}.

One can further examine Eq. (\ref{eq:sg2}) in more detail. It might be interesting to see the effect of additional term due to the interaction with the fluid represented by non-zero $U$. Keeping $U$ to be constant, \ie no viscous effect, is crucial to obtain an analytic solution. Performing a standard mathematical procedure for the traveling wave, \ie $\Phi(x,t)=\Phi(x-Ct)$ where $C$ is phase velocity, the solution is, 
\be
 \Phi(x,t) = \sqrt{\frac{2 \gamma}{\delta}} \, \mathrm{sech} \left[ \sqrt{\frac{\gamma}{C^2 - 1}} (x-Ct) \right] \; ,
 \label{eq:soliton1}
\ee 
where $\gamma \equiv m^2 - g^2 U_0^2$ and $\delta \equiv {\lambda}/{3!}$. This result is depicted in Fig. \ref{fig:soliton} for non-zero $U \equiv U_0$.

The figure depicts the amplitude as a function of $x - Ct$. It is clear that the amplitude of solitonic DNA will be damped by its interaction with surrounding fluid. In other word, the soliton with free medium ($U_0 = 0$) propagates faster than viscous medium. The result proves that one can explain the nonlinear wave equation describing the large amplitude of DNA using the Navier-Stokes lagrangian from first principle without any ambiguities due to some phenomenological considerations as done in conventional approaches. This approach is also easier than the conventional ones as for instance the derivation of nonlinear wave equation through the Navier-Stokes equation using perturbation theory.

The author greatly acknowledges L.T. Handoko for stimulation support throughout the work and reading carefully the manuscript.

\end{document}